\documentclass[prl,twocolumn,showpacs]{revtex4}

\input{epsf}
\newcommand{\bcen}{\begin{center}}
\newcommand{\ecen}{\end{center}}
\newcommand{\btab}{\begin{tabular}}
\newcommand{\etab}{\end{tabular}}
\newcommand{\bdes}{\begin{description}}
\newcommand{\edes}{\end{description}}

\newcommand{\beq}{\begin{equation}}
\newcommand{\eeq}{\end{equation}}
\newcommand{\bea}{\begin{eqnarray}}
\newcommand{\eea}{\end{eqnarray}}
\newcommand{\non}{\nonumber}

\newcommand{\bary}{\begin{array}}
\newcommand{\eary}{\end{array}}
\usepackage{epsfig}

\begin{document}
\bibliographystyle{apsrev}


\title{Templeting of Thin Films Induced by Dewetting on Patterned Surfaces}



\author{Kajari Kargupta and Ashutosh Sharma$^\ast$}
\email[]{ashutos@iitk.ac.in}
\affiliation{Department of Chemical
Engineering, Indian Institute of Technology, Kanpur, UP 208 016, India }
\date{\today}

\begin{abstract}

The instability, dynamics and morphological transitions of patterns in thin liquid films 
on periodic striped surfaces (consisting of alternating less and more wettable stripes) are  
investigated based on 3-D nonlinear simulations that account for the inter-site hydrodynamic and 
surface-energetic interactions. The film breakup is suppressed on some potentially destabilizing 
nonwettable sites when their spacing is below a characteristic lengthscale of the instability ($\lambda_h$), 
the upper bound for which is close to the spinodal lengthscale. The thin film 
pattern replicates the substrate surface energy pattern closely only when, (a) the periodicity of 
substrate pattern matches closely with the $\lambda_h$, and (b) the stripe-width is within 
a range bounded by a lower critical length, below which no heterogeneous rupture occurs, and an 
upper transition length above which complex morphological features bearing little resemblance to 
the substrate pattern are formed. 
\end{abstract}


\maketitle

Self-organization during dewetting of thin 
films on deliberately tailored chemically 
heterogeneous substrates is of 
increasing promise for engineering of 
desired nano- and micro-patterns in thin 
films by 
templeting\cite{Lenz,Gau,Kumar,Rockf,Bolt,Nisato,Gleiche,Kataoka,Hlggins,Karim,Lopez}.   
On a chemically heterogeneous substrate, dewetting is driven 
by the spatial gradient of micro-scale 
wettability \cite{Kon}, rather than by the non-
wettability of the substrate itself. The latter occurs in 
the so called spinodal dewetting on homogeneous 
surfaces \cite{Reiter,Sharma}. While the rupture of a thin film on 
a single heterogeneous patch is now well 
understood, patterned substrates pack a large 
density of surface features that are closely 
spaced. How does hydrodynamic 
interactions between the neighboring 
heterogeneities affect the pattern evolution 
dynamics and morphology in thin films? 
This question, which is addressed here, is 
central to our understanding of how 
faithfully the substrate patterns are 
reproduced in a thin film spontaneously, {\em i.e.}, 
how effective is the templeting of soft 
materials by dewetting route and what are 
the conditions for ideal templeting? An 
associated question for both the patterned 
and naturally occurring heterogeneous 
surfaces is whether all the potentially 
dewetting sites remain active or ``live'' in 
producing rupture when they are in close 
proximity. These questions are resolved 
based on 3-D nonlinear simulations of the 
stability, dynamics and morphology of thin 
films on periodic chemically heterogeneous 
surfaces. 

The substrate considered consists 
of alternating less wettable and more 
wettable (or completely wettable) stripes that differ in their 
interactions with the overlying film. The key 
parameters of the substrate pattern are its 
periodicity interval (center-to-center distance between two 
consecutive stripes, $L_p$) and the length-scale 
of the less wettable stripe (stripe-width, $W$).  
The following nondimensional thin film 
equation governs the stability and spatio-
temporal evolution of a thin film system 
subjected to the excess intermolecular 
interactions \cite{Kon}. 
\begin{equation}
\partial H / \partial T + \nabla \cdot  [H^3 \nabla (\nabla ^ 2 H)] -
\nabla \cdot [H^3 \nabla\Phi] = 0
\end{equation}
H(X,Y,T) is  non-dimensional local film 
thickness scaled with the mean thickness $h_o$; 
$\Phi= [2\pi h_o^2/|A_s|][\partial \Delta G/\partial H])$;
$\Delta G$ is the excess intermolecular interaction energy per unit area and 
$A_s$ is the effective Hamaker constant for van der Waals interaction; 
$X$, $Y$ are the non-dimensional coordinates in the plane of
the substrate, scaled with a lengthscale  
$(2 \pi \gamma / |A_s|)^{1/2} h_o^2$; and the non-dimensional time $T$ is scaled
with $12 \pi ^2 \mu \gamma h_o^5 / A_s^2$; 
$\gamma$ and $\mu$ refer to the film surface tension and
viscosity, respectively. 
The lengthscale of the 
spinodal instability on a uniform surface is 
given by, 
$\lambda=(-4\pi^2 \gamma/(\partial^2 \Delta G /\partial h^2))^{1/2}$. 
On a chemically heterogeneous striped 
surface, $\Phi = \Phi(H, X, Y)$. At a constant film 
thickness, we model the variation of $\Phi$ in X 
direction by a periodic step function of 
periodicity, $L_p$. Gradient of force $\nabla\Phi$, at the 
boundary of the stripes causes flow from the 
less wettable (higher pressure) stripes to the 
more wettable (lower pressure) stripes, even 
when the spinodal stability condition $\partial \Phi / \partial H>0$ 
is satisfied everywhere \cite{Kon}. It is 
known that a single stripe in the absence of 
its neighbors can cause rupture only if its 
width exceeds a critical lengthscale, $W_c << \lambda$
 \cite{Kon}.    

We consider general representation 
of antagonistic (attractive/repulsive) long 
range and short range intermolecular 
interactions applicable to the aqueous films (eq. 2) \cite{Thiele,Sharma} and polymer films on high energy surfaces like sillicon (eq. 3) \cite{Kim,Kon,Rockf}. 
In both the cases the long range van der Waals force is stabilizing and thus films thinner (thicker) than a critical thickness are unstable (metastable).
\begin{equation}
\Delta G = -(A_s/12\pi h^2)
+ S_p \exp(-h/l_p)
\end{equation}
This potential represents   
aqueous films when the effective 
Hamaker constant, $A_s$ and $S_p$ are both 
negative (long-range van der Waals 
repulsion combined with shorter-range  
attraction, e.g. hydrophobic attraction)\cite{Van}. On the hydrophobic stripes ($A_s^h 
= -1.41$ x $10^{-20}$ J, $S_p^h = -65$ mJ/$m^2$ and $l_p$ = 
0.6 nm), the film is spinodally unstable 
($\partial \Phi/ \partial H<0$) below a critical thickness, $h_c$ of 
7.4 nm. The equilibrium contact angle 
obtained from extended Young-Dupre 
equation, $cos\theta = 1+\Delta G(h_e)/\gamma$ is $61^o$; where $h_e$ is the 
equilibrium thickness. On 
the hydrophilic stripes ($A_s$ = $-1.41$ x $10^{-20}$ J, 
$S_p$ = $-0.61$ mJ/$m^2$ and $l_p$ = 0.6 nm), the film 
of any thickness is spinodally stable  
($\partial \Phi/ \partial H>0$) 
and completely wets the surface. 

Equation (1) was numerically solved using a 
central difference scheme in space combined 
with Gears algorithm for stiff equations for 
time marching and periodic boundary 
condition. The critical length, $W_c$ of a single 
heterogeneity (hydrophobic patch) that 
engenders rupture on a large hydrophilic 
substrate in this case is $0.08\lambda$. 

\begin{figure}
\vspace{-5.9truecm}
\centerline{\psfig{file=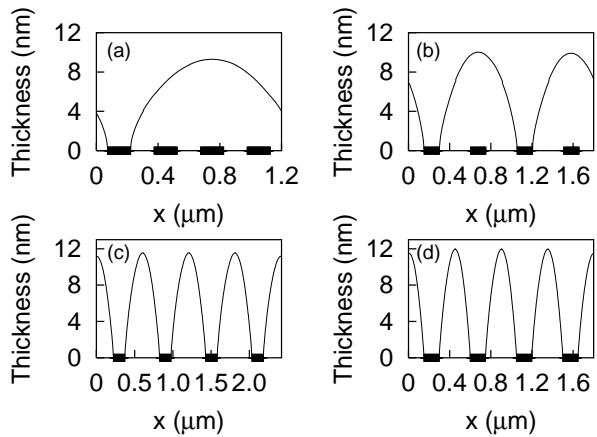,width=3.30in}}
\vspace{-0.9truecm}
\caption {Equilibrium thin film profile on a 
striped surface ($W$ = 0.16$\lambda$). The less-wettable 
stripes are denoted by black rectangles.  Figs 1a, 
1b and 1c is for a 6nm film for increasing 
periodicity. Fig 1d shows the profile for a 5.5 nm 
thin film on a same substrate shown in Fig. 1b.}  
\end{figure}
Ideal templating requires that dewetting should occur on every less wettable site on the patterned substrate.
Fig. 1a shows a single rupture on a substrate 
containing four potentially dewetting sites 
($n_S$ = 4) when the periodicity interval is 
small, $L_p$ = $0.45\lambda$. The resulting lone drop 
spans across the remaining ``inactive''
hydrophobic sites. Fig 1a to 1c show 
increased number of dewetted stripes ($n_D$ = 1, 
2, 4) and resulting liquid domains as $L_p$ is 
increased from $0.45\lambda$ $\rightarrow$  $0.68\lambda$ $\rightarrow$ $0.9\lambda$. 
This, as well as a large number of 
simulations not displayed here showed that 
on a substrate containing many potentially 
destabilizing sites, only the sites (randomly 
picked) separated by a characteristic lengthscale of the instability, $\lambda_h$, of the order 
of the spinodal scale, $\lambda$ remain ``live'' or 
effective in causing the film rupture.  
Dewetting on the remaining intervening 
sites is suppressed since rupture on each site 
would require surface deformations on 
smaller scales resulting in high surface 
energy penalty. The ratio of this characteristic lengthscale, $\lambda_h$, to the spinodal length scale, $\lambda$, obtained from simulations, decreases as the potential difference across the stripe bounadry, ($\Delta \Phi$) increases (Fig. 2), where $\Delta \Phi= (\Phi^h - \Phi$), evaluated at the initial thickenss. For a large value of spinodal parameter (e.g. for thinner films), the ratio approaches very close to 1 unless $\Delta \Phi$ is very large (curves 2 and 3 of Fig. 2). A simple scale analysis ($H=  1 + \epsilon$, $\nabla^2 \Phi \sim \Delta \Phi/cL^2$, $ c< 1$), of  
Eq. (1) \cite{Kon} leads 
to the following characteristic lengthscale for growth of instability:
$\lambda_h \propto  [\Delta \Phi /\epsilon c - \partial \Phi/\partial H]^{-0.5}$.
Thus the ratio of heterogeneous to spinodal lengthscale, $\lambda_h/\lambda \propto (1 + \Delta \Phi/(-
\epsilon c  \partial \Phi/\partial H))^{-0.5}$; and  $\lambda_h \sim \lambda$
when spinodal term $-\partial \Phi/\partial H$ is very strong compared to the applied 
potential difference.  
\begin{figure}
\vspace{-5truecm}
\centerline{\psfig{file=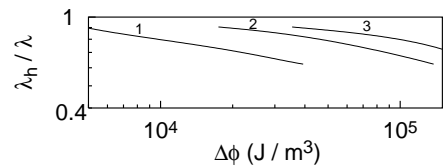,width=3.0in}}
\vspace{-4.truecm}
\caption { Variation of the ratio of characteristic to spinodal length scale with potential difference 
$\Delta\phi$, introduced by changing the $S_p$ of the more wettable stripes. $-\partial\phi/\partial 
h$, for curves 1 ($h_o =6nm$, $S_p^h=-65mJ/m^2$), 2 ($h_o= 
4.5nm$, $S_p^h= -19.5mJ/m^2$) and 3 ($h_o=4.5 nm$, $S_p^h= -39mJ/m^2$) 
are $6.47$x$10^{12}$, $2.4$x$10^{13}$ and $5.4$x$10^{13}$ $J / m^4$, respectively. } 
\vspace{-0.5truecm}
\end{figure}
For very large potential difference, $\lambda_h/\lambda \propto (\Delta\phi)^{-0.5}$.
Thus the upper limit of the characteristic lengthscale of the instability is always bound by the spinodal length, $\lambda$.   
An interesting implication is 
that even when the rupture occurs by the 
heterogeneous mechanism on a substrate 
containing a large density of heterogeneities, 
the lengthscale of the resulting pattern can 
give the illusion of spinodal dewetting by 
mimicking the characteristic lengthscale of 
the latter. Differentiating true spinodal 
dewetting from heterogeneous dewetting 
therefore requires a careful consideration of 
their distinct time scales and 
morphological features\cite{Kon}. Further, the 
contact line always remains pinned at the 
boundary of the stripes whenever alternate 
stripes are completely wettable (Figures 1a 
$-$ 1d).

Decreased mean film thickness increases both the terms, $\Delta \Phi$ 
and $-\partial \Phi/\partial h$ and thus decreases  $\lambda_h$ nonlineraly.      
Thus, a greater number of 
nonwettable heterogeneities become active 
in causing the film breakup for thinner films. 
Comparison of figs. 1b (thickness = 6 nm, $\lambda$  
= 0.67 $\mu$m) and 1d (thickness = 5.5 nm, $\lambda$ = 
 0.42 $\mu$m) clearly shows this for an identical 
substrate pattern. This transition uncovered 
based on dynamical simulations is also in 
conformity with the earlier equilibrium 
energy considerations \cite{Lenz} that are related however only to the equilibrium structures. 
 
From the above discussion, a necessary condition for good templating is that $L_p \sim \lambda_h (< \lambda)$ which ensures dewetting on every less wettable site ($n_D = n_S$).
The condition $L_p$ $\sim$ $\lambda_h$, for $n_D = n_S$, also 
remains valid when both the stripes are 
nonwettable. However, in this case the 
contact line can move across the boundary 
of the stripes to the more wettable part.
We simulated the 3D morphologies during the 
evolution of a polymer like film on an oxide 
(low-energy) covered silicon (high-energy) 
substrate. An analytical 
representation of combined antagonistic 
(attractive/repulsive) long and short-range 
intermolecular interactions for a polymer-like film on a 
coated (e.g. oxide covered) substrate is\cite{Kon}:
{\small 
\bea
-12\pi \Delta G = [(A_s - A_{c1})/(h+d_{c1}+d_{c2})^2 
\non
\\
 +  (A_{c1} - A_{c2})/(h+d_{c2})^{2} + A_{c2}/h^2]
\eea
}
Negative value of the effective Hamaker 
constant on the substrate [$A_s =  A_s^h =  
-1.88$ x $10^{-20}$ J] signifies a long-range 
repulsion, whereas a positive value on 
coating [$A_{c1}= A_{c1}^h = 1.13$ x $10^{-20}$ J] represents 
an intermediate-range attraction\cite{Kon}. The 
non-wettable coating (e.g. oxide) thickness 
[$d_{c1} = 2.5$ nm, $d_{c1}^h = 4$ nm] is increased on 
alternating stripes which causes the 
macroscopic contact angle to increase from 
$0.58^o$ on the more wettable stripes to $1.7^o$ on 
the less wettable stripes.  A still shorter-
range repulsion may arise due to chemically 
adsorbed or grafted layer of the polymer 
($A_{c2} =  -0.188$ x $10^{-20}$ J, $d_{c2}  = 1$nm are the 
film Hamaker constant and the thickness of 
the adsorbed layer, respectively)\cite{Kon}. The 
representation, eq.[3], is chosen for 
illustration without affecting the underlying 
physics, which we have verified for many 
different potentials.
\begin{figure}
\vspace{1.8truecm}
\centerline{\psfig{file=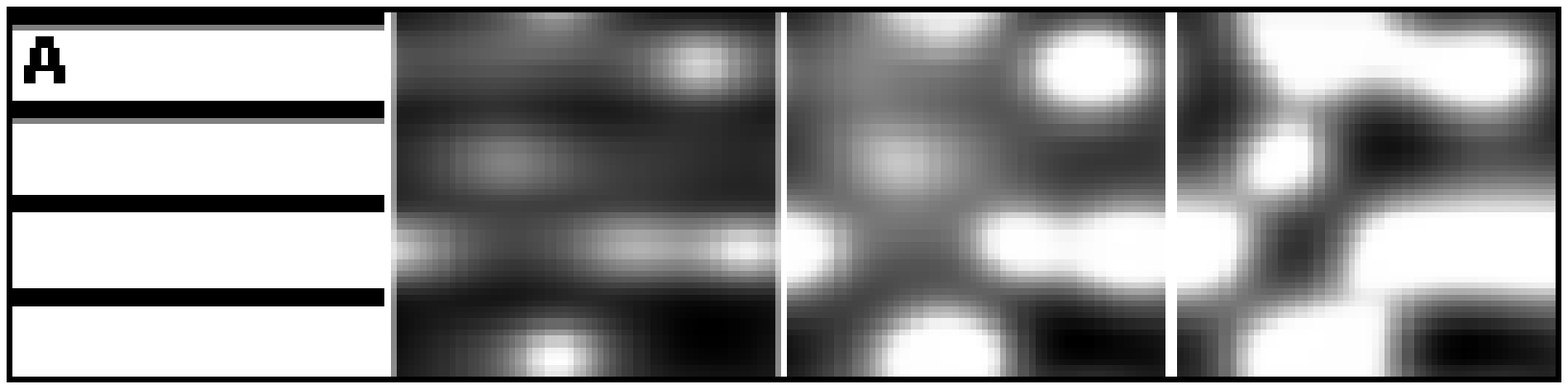,width=3.1in}}
\vspace{1.8truecm}
\centerline{\psfig{file=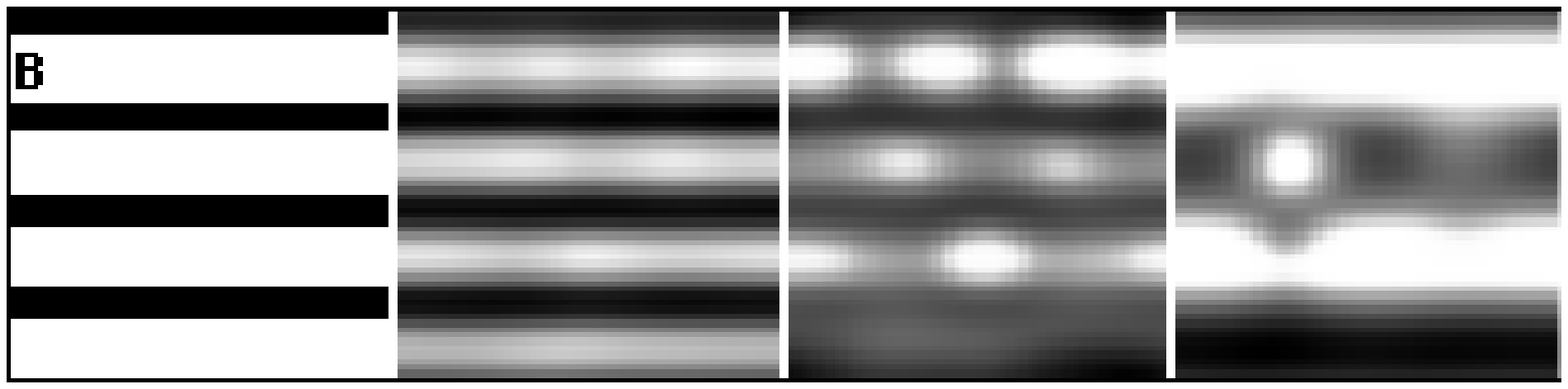,width=3.1in}}
\vspace{1.8truecm}
\centerline{\psfig{file=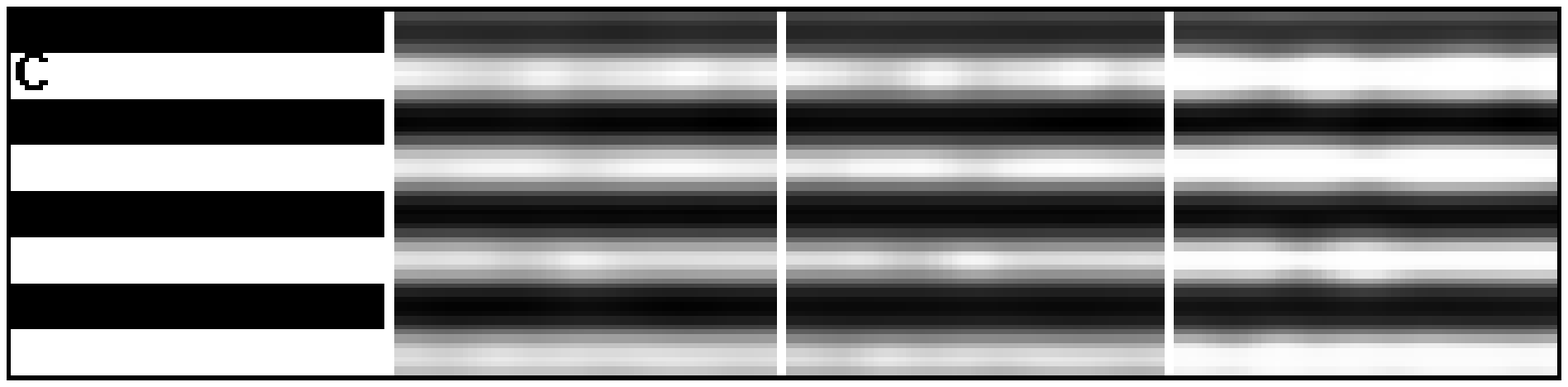,width=3.10in}}
\vspace{1.8truecm}
\centerline{\psfig{file=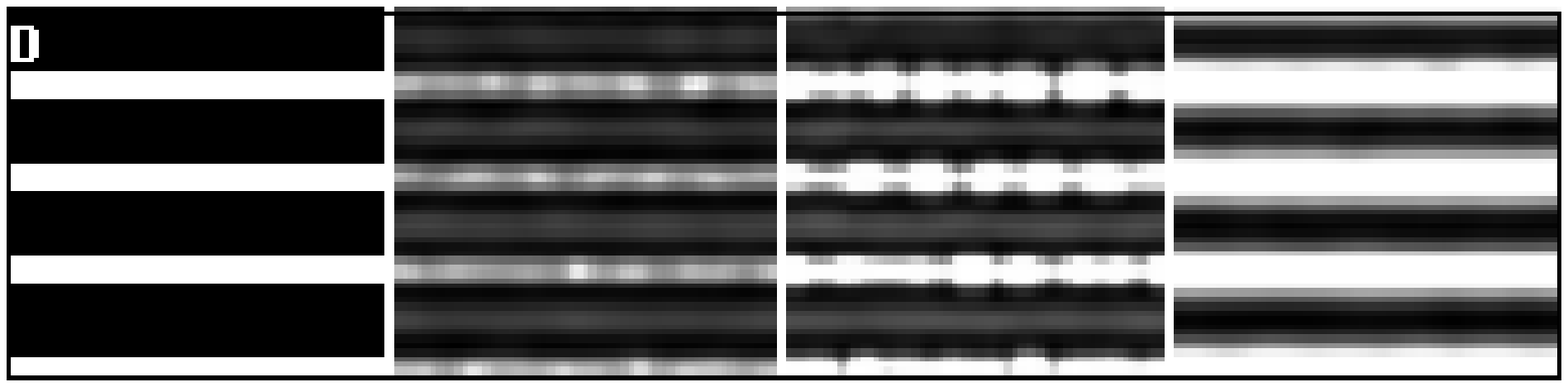,width=3.10in}}
\caption {Morphological evolution in a 5 nm thick 
film on a striped surface ($W$ = 0.8 $\mu$m = 0.53 $\lambda$) 
In Figures 3A to 3D periodicity of surface pattern, 
$L_p$ are 1 $\mu$m, 1.2 $\mu$m, 1.45 $\mu$m and 3 $\mu$m 
respectively. The first image in this figure as 
well as in the subsequent figures represents the 
substrate surface energy pattern; black and white 
represent the more wettable part and less 
wettable part respectively. For other images 
describing the film morphology, a continuous 
linear gray scale between the minimum and the 
maximum thickness in each picture has been 
used.} 
\end{figure}

Figures 3A $-$ 3D depict the effect of 
periodicity interval, $L_p$ on the self-
organization of a 5 nm thin film that is 
spinodally unstable on both type of stripes. 
In each case, the evolution starts with local 
depressions on the less-wettable stripes 
(Figs. 3A $-$ 3D). For $L_p$ sufficiently smaller 
than $\lambda$($L_p = 1 \mu$m, $\lambda = 1.51  \mu$m based on $\gamma 
= 38$ mJ/$m^2$ and $\mu = 1  kg/m.s$), the isolated 
holes or depressions that form along the less wettable 
stripes grow onto the more wettable regions 
and coalesce with each other rapidly (Fig. 
3A) leading to a disordered structure and 
very poor templeting. Increasing periodicity 
($L_p = 1.2 \mu$m; Fig. 3B) leads to more ordered 
dewetting, but the number of dewetted 
regions remain less than the number of less 
wettable stripes, 
and defects evolve at late times (e.g., holes 
in the liquid ridge; image 4 of fig. 3B). For 
$L_p \geq \lambda$ (Figures 3C and 3D), the number of 
dewetted stripes equal the number of less 
wettable stripes. However, templeting in the 
form of liquid ridges with straight edges is 
best at an intermediate periodicity, $L_p \sim \lambda_h (< \lambda)$ 
(Fig. 3C). A further increase in $L_p$ makes the 
width of dewetted region bigger than the 
width of the less wettable stripe (image 4 of 
Fig. 3D), {\em i.e.}, the contact line resides in the 
interior of the more wettable stripe, rather 
than close to the boundary. 
\begin{figure}
\vspace{1.8truecm}
\centerline{\psfig{file=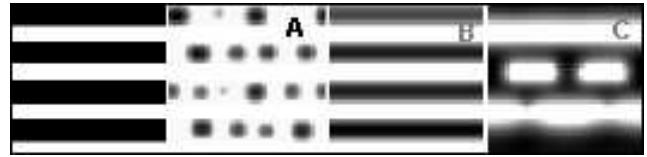,width=3.1in}}
\caption {3D morphologies at late stages of evolution 
for films of thickness (A) 2.5 nm, (B) 6.9 nm and 
(C) 8 nm on a striped surface with $W$ = 1.2 $\mu$m 
and $L_p$ = 3 $\mu$m.}
\end{figure}

The transition of surface morphology can 
also be clearly understood from Fig. 4, 
which shows that ideal templeting occurs for 
an intermediate thickness (3rd image of Fig. 
4) for which the spinodal wavelength, $\lambda(= 3 
\mu$m) is equal to $L_p$. Thinner films evolve by 
the formation of droplets on the more 
wettable stripes due to the spinodal 
mechanism assisted by the Rayleigh 
instability. For thicker films  ($\lambda > L_p$), 
dewetting occurs on fewer stripes with the 
formation of holes in broad liquid ridges 
(e.g., last image of Fig.4).  

We found that besides $L_p$, the other 
parameter that governs the thin film pattern 
is the stripe-width ($W$). It is already known 
that for a single heterogeneity on a large 
substrate, increase in the width of the 
heterogeneity shifts the onset of dewetting 
from the center of the heterogeneity to the 
boundary of the heterogeneity \cite{Kon}. Figs. 
\begin{figure}
\vspace{1.8truecm}
\centerline{\psfig{file=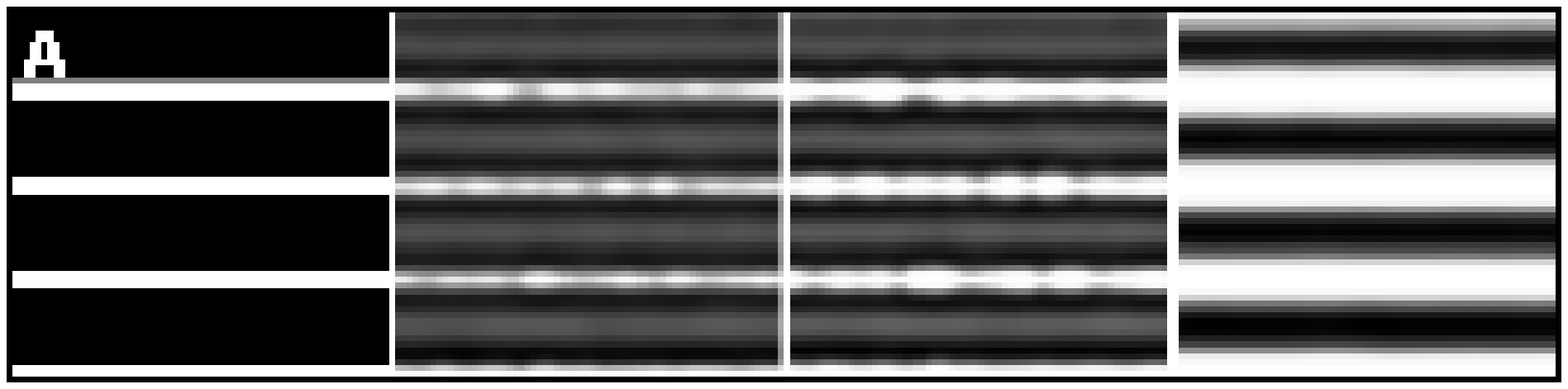,width=3.1in}}
\vspace{1.8truecm}
\centerline{\psfig{file=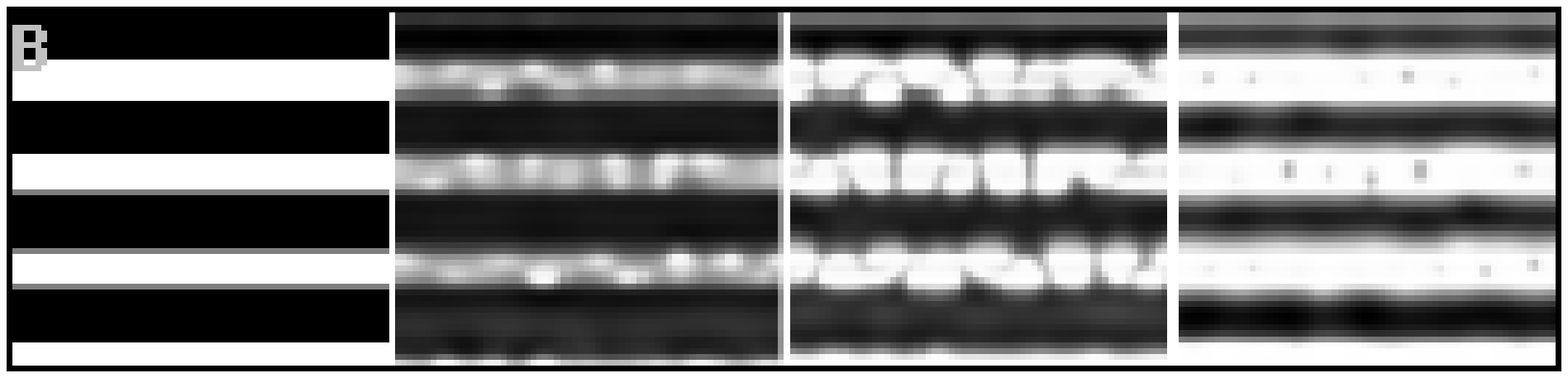,width=3.1in}}
\vspace{1.8truecm}
\centerline{\psfig{file=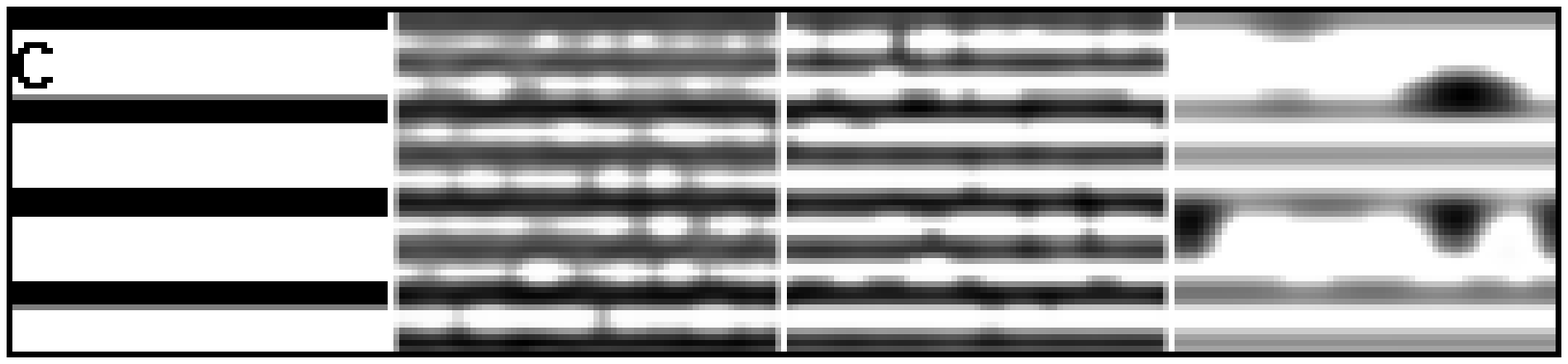,width=3.1in}}
\vspace{1.8truecm}
\centerline{\psfig{file=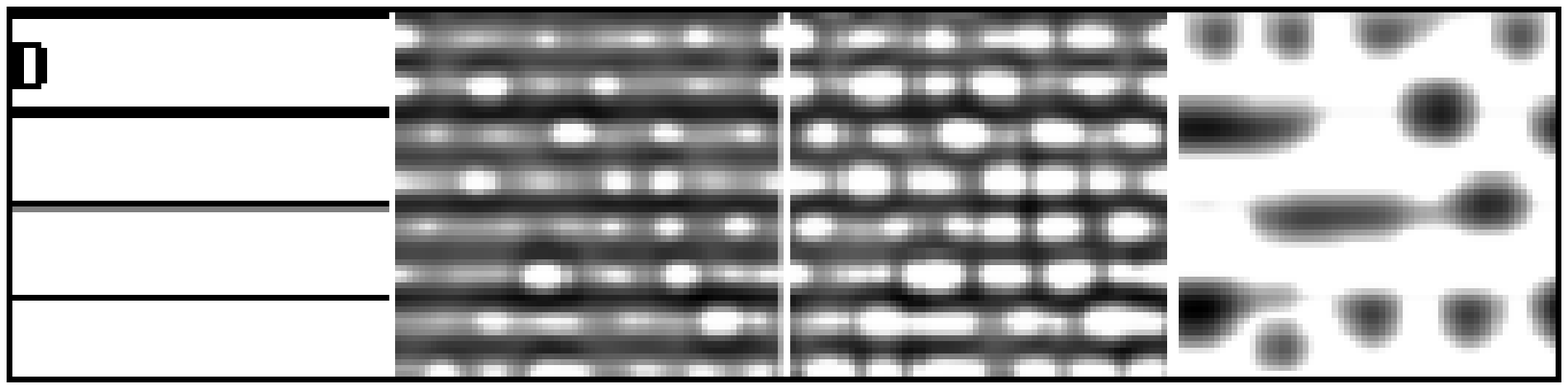,width=3.1in}}
\caption {Morphological evolution in a 5 nm thick 
film on a striped surface ($L_p$ = 3 $\mu$m $\sim 2 \lambda$).    
$W$ =  0.6 $\mu$m, 1.2 $\mu$ 
m, 2.1 $\mu$m and 2.7 $\mu$m, respectively for figures 5A $-$ 5D.}
\end{figure}
5A $-$ 5D shows the transition of patterning 
with the increase of the stripe-width ($W$), 
keeping the periodicity, $L_p$ fixed ($> \lambda$). For 
small stripe-width ($\sim 0.4 \lambda$; Fig. 5A), rupture 
is initiated at the center of less wettable 
stripes by the formation of depressions that 
coalesce to form rectangular dewetted 
regions of widths greater than the stripe-
width. For a larger stripe-width ($W \sim 0.8\lambda$), 
dewetting is initiated by a layer of holes at 
each of the two boundaries of the stripe 
(image 2 of Fig. 5B). Coalescence of these 
two layers of holes leads to the dewetted 
regions containing some residual droplets 
(image 4 of Fig. 5B). Further increase in the 
stripe width leads to the formation of two 
layers of holes on each stripe separated by 
an elevated liquid cylinder (image 2 of 
Fig.5C). Thus, at an intermediate stage of 
evolution, the number of cylindrical liquid 
ridges becomes twice of the number of more 
wettable stripes on the substrate (image 3 of 
Fig. 5C). Eventually, liquid ridges on the less 
wettable region disintegrate into droplets 
(not shown). Laplace pressure gradients 
cause ripening of the structure leading to the 
merging of droplets with the liquid ridges 
(image 4 of Fig. 5C). Further increase in the 
stripe-width breaks the whole order of the 
substrate pattern due to the formation and 
repeated coalescence of several layers of 
holes on each stripe that eventually evolve 
into arrays of irregular droplets. Thus, a 
good synchronization of the thin film 
morphology with the substrate pattern 
requires an upper limit on the stripe-width 
($W_t \sim < 0.8\lambda$, for this case). Based on a large 
number of simulations for different film 
thickness and for different systems, we have 
verified that a breakdown of templeting due 
to the off-center dewetting occurs at stripe-
widths in excess of about $0.7\lambda - 0.8\lambda$. 

In conclusion, the 3-D thin film morphology 
on a patterned substrate during dewetting 
can be modulated profoundly by a 
competition among the time scales (spinodal 
and heterogeneous) and length scales 
(spinodal, stripe-width, periodicity, 
thickness) of the problem. On a striped 
substrate, ideal templeting in the form of 
cylindrical liquid ridges covering the more 
wettable stripes occurs when (e.g., image 4 
in fig. 3C): (a) periodicity of the substrate 
pattern is very close to the characteristic lengthscale, $\lambda_h$ 
, and (b) the stripe width is 
larger than a critical width, $W_c$, which is 
effective in causing rupture by the 
heterogeneous mechanism, but smaller than 
a transition width ($W_t \sim 0.7\lambda - 0.8\lambda$), that 
ensures initiation of dewetting at the stripe-
center. Even a very weak wettability contrast (e.g. Fig.3) is sufficient to align the thin film pattern with the templete under the above conditions. Predictions of our simulations show 
close resemblance to the recently reported 
experimental observations on dewetting of  
polymers on  patterned surfaces that reveal: 
(a) best templeting for an intermediate 
thickness film\cite{Nisato,Karim} and (b) correspondence 
between the natural length scale and 
periodicity of the substrate pattern for good 
templeting\cite{Rockf,Bolt,Nisato}. Finally, on a completely 
wettable substrate containing many 
potentially destabilizing sites, only the sites 
(randomly picked) separated by a characteristic lengthscale (of 
the order of the spinodal scale), $\lambda_h$ remain 
``live'' or effective in causing the film 
rupture. It is hoped that this study will help 
in the design and interpretation, creation and rational manipulation of self organized microstructures in thin films by templeting.  

\vspace{0.1truecm}

\begin{acknowledgments}
Discussions with G. Reiter are  gratefully acknowledged. 
\end{acknowledgments}

\medskip
\hrule

\vspace{0.5truecm}
\bibliography{stripe}
~
~
~
~
~

\end{document}